# Phase-encoded RF signal generation based on an integrated 49GHz micro-comb optical source

Xingyuan Xu, Mengxi Tan, Jiayang Wu, Andreas Boes, Bill Corcoran, Thach G. Nguyen, Sai T. Chu, Brent E. Little, Roberto Morandotti, Arnan Mitchell, *and* David J. Moss

***Abstract*—We demonstrate photonic RF phase encoding based on an integrated micro-comb source. By assembling single-cycle Gaussian pulse replicas using a transversal filtering structure, phase encoded waveforms can be generated by programming the weights of the wavelength channels. This approach eliminates the need for RF signal generators for RF carrier generation or arbitrary waveform generators for phase encoded signal generation. A large number of wavelengths—up to 60—were provided by the microcomb source, yielding a high pulse compression ratio of 30. Reconfigurable phase encoding rates ranging from 2 to 6 Gb/s were achieved by adjusting the length of each phase code. This work demonstrates the significant potentials of this microcomb-based approach to achieve high-speed RF photonic phase encoding with low cost and footprint.**

## I. INTRODUCTION

Phase encoded continuous wave (CW) radio frequency (RF) signals are widely employed in low probability of intercept (LPI) radar systems, since they feature low power density and employ random codes to avoid being cracked [1-2]. In such systems, targets are resolved within range when the reflected phase encoded signals are compressed at the receiver, and a large signal bandwidth is required to achieve a high resolution [3]. Photonic techniques have significant potential to overcome the limitations of their electronic counterparts [4] due to their intrinsically broad bandwidth, immunity from electromagnetic interference, and low propagation loss [5-11].

Significant effort has been devoted to realizing photonic-assisted RF phase encoding, including approaches based on polarization modulators [12, 13], Sagnac loops [14, 15], and dual parallel modulators [16, 17]. However, these approaches all face limitations of one form or another, such as the need for complicated modulation schemes to achieve the required π-phase shifts, and the need for both high-frequency RF signal generators and arbitrary waveform generators. The former increases the complexity and instability (such as the bias drift of the modulators) of the system, while the latter greatly increase the cost, size and power consumption.

An alternative method of generating RF phase encoded signals based on transversal filtering structures has been proposed [18]. In this method, RF high-order Gaussian pulses are broadcast onto 4 wavelength channels via an intensity modulator, and then progressively delayed through a dispersive medium such that the pulse segments are assembled (concatenated) in the time domain. By controlling the phase of each pulse segment, phase-encoded waveforms can be generated. This approach reduces the need for RF sources and complicated modulation schemes, and offers a high potential phase encoding speed by tailoring the delay step and pulse width. However, it is not without challenges. The use of discrete laser arrays significantly increases the complexity and cost for further integration and limits the number of wavelengths. This in turn leads to a limited sequence length and thus a small pulse compression ratio that would limit the performance of radar systems. Secondly, the employed high-order Gaussian pulses require complicated

---

This work was supported by the Australian Research Council Discovery Projects Program (No. DP150104327). RM acknowledges support by the Natural Sciences and Engineering Research Council of Canada (NSERC) through the Strategic, Discovery and Acceleration Grants Schemes, by the MESI PSR-SIIRI Initiative in Quebec, and by the Canada Research Chair Program. He also acknowledges additional support by the 1000 Talents Sichuan Program in China. Brent E. Little was supported by the Strategic Priority Research Program of the Chinese Academy of Sciences, Grant No. XDB24030000.

X. Xu, M. Tan, J. Wu, and D. J. Moss are with the Centre for Micro-Photonics, Swinburne University of Technology, Hawthorn, VIC 3122, Australia. (Corresponding e-mail: dmoss@swin.edu.au).

A. Boes, T. G. Thach and A. Mitchell are with the School of Engineering, RMIT University, Melbourne, VIC 3001, Australia.

B. Corcoran is with the Department of Electrical and Computer System Engineering, Monash University, Clayton, VIC 3168 Australia.

S. T. Chu is with the Department of Physics, City University of Hong Kong, Tat Chee Avenue, Hong Kong, China.

B. E. Little is with State Key Laboratory of Transient Optics and Photonics, Xi'an Institute of Optics and Precision Mechanics, Chinese Academy of Science, Xi'an, China.

R. Morandotti is with INSR-Énergie, Matériaux et Télécommunications, 1650 Boulevard Lionel-Boulet, Varennes, Québec, J3X 1S2, Canada, and adjunct with the Institute of Fundamental and Frontier Sciences, University of Electronic Science and Technology of China, Chengdu 610054, China.

generation schemes such as high-order differentiation, which will increase the cost and complexity of the system.

Integrated microcombs [19-24], generated through optical parametric oscillation in monolithic microring resonators (MRRs), are promising multi-wavelength sources that offer many advantages including a much higher number of wavelengths and greatly reduced footprint and complexity. Thus, a wide range of RF applications have been demonstrated based on micro-combs, such as transversal filters [25-27], differentiators [28], optical true time delays [29, 30] and channelizers [31, 32], Hilbert transformers [33-35], microwave mixers [36], microwave generation [37], and more.

In this paper, we demonstrate phase encoded CW RF signal generation using an integrated soliton crystal micro-comb source. The soliton crystal is a recently discovered coherent and stable solution of the Lugiato-Lefever equation [38, 39], which governs the nonlinear dynamics of the optical parametric oscillator. The soliton crystal stably forms through the background wave generated by a mode crossing, and features over thirty times higher intra-cavity power than traditional dissipative Kerr solitons. Thus, they can remove thermal effects and enable adiabatic pump sweeping for coherent comb generation [39, 40]. These properties, together with the broadband comb spectrum, make soliton crystals promising for demanding integrated RF photonic systems, such as the photonic RF phase-encoder introduced in this work.

Here, by multicasting an RF single-cycle pulse (instead of the high-order pulse used in [18]) onto a spectrally flattened micro-comb and progressively delaying the replicas with a dispersive medium, the RF pulse replicas are assembled arbitrarily in time according to the designed binary phase codes. The large number of wavelengths offered by the microcomb enabled 30 bits, representing an enhancement of over 4 times compared to previous work [18], for the phased encoded RF sequence, leading to a pulse compression ratio as high as 29.6. A reconfigurable phase encoding rate ranging from 1.98 to 5.95 Gb/s was also achieved by varying the length of each phase code. These results verify the strong potential of our approach for the realization of photonic RF phase encoding schemes and represent a solid step towards the miniaturization of fully integrated radar transmitters with greatly reduced cost and footprint.

## II. THEORY

Figure 1 illustrates the operation principle of a photonic phase encoder. First, an RF Gaussian pulse $f(t)$ with a duration of $\Delta t$ is generated. Then a discrete convolution operation between the RF pulse and flattened microcomb spectrum (denoted by a discrete signal $g[n]$ with length $N$ and binary values of 1 or -1) is performed with a delay step of $\Delta t$, and can be described as:

$$(f * g)[n] = \sum_{i=1}^{N} f[n - i \cdot \Delta t] \cdot g[i] \tag{2.1}$$

The discrete convolution operation between the RF pulse $f(t)$ and the flattened microcomb spectrum is achieved using the experimental setup in Figure 2. The RF pulse is first multi-cast onto the wavelength channels to yield replicas, which are then delayed with a progressive step that matches the pulse duration $\Delta t$. By flipping the sign of the delayed replicas according to designed phase codes $g[n]$, a phase-encoded sequence can be assembled in the time domain. The sign reversal is accomplished by spatially separating the intended positive pulses from the negative ones using the programmable wavelength dependent spatial routing capability of a waveshaper, and then directing the two separate outputs to the positive and negative inputs of a differential photodetector, respectively. The total number of RF pulse replicas $N$ is equal to the number of wavelength channels, which is 60 in our case—15 times that of the previous work [18]. Thus, the total time length of the phase-encoded sequence is $T=N\cdot\Delta t$. Here, the basic temporal element in the phase-encoded sequence is termed an "RF segment", which is a single-cycle or multi-cycle sine wave assembled from the input RF pulse (i.e., two anti-phase RF pulses assemble a single-cycle sine wave). The centre frequency of the phase-encoded sequence is determined by the frequency of the assembled sine waves, thus is equivalently given by $1/(2\Delta t)$. Assuming each RF segment constitutes $m$ pulses, then the length of the phase-encoded sequence would be $N/m$, together with an equivalent phase encoding speed of $1/(m\,\Delta t)$. We note that the generated phase-encoded sequence has a bandwidth similar to the input RF pulse, which is subject to the Nyquist bandwidth, equivalent to half of the comb spacing (48.9 GHz/2=24.45 GHz).

In a radar system, the phase encoded RF waveform is first amplified and then emitted by antennas, followed by reflection from the target, and finally collected and processed at the receiver. The range information of the target is carried on the delay of the received waveforms, which can be readily determined by calculating the autocorrelation function of the reflected phase-encoded RF signal given by

$$ACF = \int_{-\infty}^{+\infty} s(\tau)s(\tau+t)d\tau \qquad (2.2)$$

In practical systems this is calculated by performing a hardware autocorrelation extracted by an array of range gates. The range gates are composed of matched filters with their tap coefficients determined by the employed phase codes (Fig. 3). The function of the matched filters is to perform single-output autocorrelations for the received signal, and the range gate that has the highest output power denotes the resolved range of the target.

## III. EXPERIMENTAL RESULTS

The experimental setup is shown in Figure 2. A CW laser is amplified and its polarization state adjusted to pump a nonlinear high Q factor (> 1.5 million) MRR, which featured a free spectral range of ~0.4nm, or 48.9 GHz. As the detuning between the pump laser and the MRR was changed, dynamic parametric oscillation states corresponding to distinctive solutions of the Lugiato-Lefever equation were initiated [38]. We thus generated soliton crystal microcombs [39, 40], which were tightly packaged solitons circulating in the MRR as a result of a mode crossing (at ~1552 nm in our case), and which resulted in the distinctive palm-like comb spectrum (Fig. 4).

Next, 60 lines of the microcomb were flattened ($N$=60 in our case), using two stages of WaveShapers (Finisar 4000S) to acquire a high link gain and signal-to-noise ratio. This was achieved by pre-flattening the microcomb lines with the first WaveShaper such that the optical power distribution of the wavelength channels roughly matched with the desired channel weights. The second WaveShaper was employed for accurate comb shaping assisted by a feedback loop as well as to separate the wavelength channels into two parts (port 1 and port 2 of the Waveshaper) according to the polarity of the designed binary phase codes. The feedback loop was constructed by reading the optical spectrum with an optical spectrum analyser and comparing the power of the wavelengths with the designed weights to generate an error signal, which was then fed back into the second WaveShaper (WS2) to calibrate its loss until the error was below 0.2 dB.

Here, we used a Gaussian pulse with a duration of $\Delta t$ = 84 ps, as the RF fragment $f[t]$. Although we used an arbitrary waveform generator (Keysight, 65 GSa/s) to generate the Gaussian pulse for the sake of simplicity, the arbitrary waveform generator is actually not a necessity and can be replaced with many other readily available approaches that are easier and cheaper [41, 42]. The input RF pulse was imprinted onto the comb lines, generating replicas across all the wavelength channels. The replicas then went through a ~13 km long spool of standard single mode fibre (with a dispersion of 17ps/nm/km) to progressively delay them, leading to a delay step of ~84 ps between the adjacent wavelength channels that matched with the duration of the RF pulse $\Delta t$. Finally, the wavelength channels were separated into two parts according to the designed phase codes and sent to a balanced photodetector (Finisar, 40 GHz) to achieve negative and positive replicas for the phase encoding.

Figure 5 shows the input Gaussian pulse and the flattened optical comb spectra measured at the output ports of the second Waveshaper (WS2). Each pair of wavelength channels from different ports can assemble a monocycle sine wave, thus with the 60 comb lines, 30 sine cycles can be achieved, with a total time length of $T = N\cdot\Delta t$ = 60×84ps = 5.04 ns.

By applying designed phase codes during the separation of the wavelength channels, the sine cycles could be π-phase shifted at desired times. The phase-encoded results are shown in Fig. 6. The number of Gaussian pulses for each RF segment (denoted by $m$) was reconfigured from 6 to 2, corresponding to reconfigurable sequence lengths ($N/m$) ranging from 10 to 30 and phase coding speeds ($1/\Delta t/m$) ranging from 1.98 to 5.95 Gb/s. The employed phase codes were denoted both by the shaded areas and the stair waveforms (black solid line). This result shows that our photonic phase coder can offer a reconfigurable sequence length to address the performance tradeoffs between range resolution and system complexity. To acquire a large pulse compression ratio for a high resolution, the sequence length should be maximized, where the number of Gaussian pulses for each RF segment ($m$) should be set as 2. Further, to reduce the complexity and cost of the RF system (such as the number of range gates at the receiver), the sequence length could be reduced by either employing fewer wavelength channels or by increasing $m$.

The corresponding optical spectra (Fig. 6 (a, c, e)) were measured at the output of Waveshaper to show the positive and negative phase codes realized by changing the wavelength channels' output ports at the Waveshaper. The encoded RF waveforms (Fig. 6 (b, d, f)) clearly show the flipped phase of the RF segments at the time of negative phase codes, where the number of sine cycles was reconfigured as well, depending on the value of $m$. This result also shows that our approach is fully reconfigurable for different phase codes and encoding speeds. We note that higher encoding speeds can be achieved by reducing the duration of the RF fragment and the delay step $\Delta t$.

In this work we calculated the autocorrelation (Fig. 7) of the phase-encoded RF waveforms $s(t)$. As the sequence length varied from 10 to 30, the full width at half-maximum (FWHM) of the compressed pulses varied from 0.52 to 0.17 ns, which corresponds to a pulse suppression ratio ranging from 9.7 to 29.6. Meanwhile the peak-to-sidelobe ratio (PSR) also increased with the sequence length from 4.17 to 6.59 dB. These results confirm that the pulse compression ratio of an RF phase-encoded signal is linearly related to its sequence length [43], and that this can be significantly enhanced with our approach by employing a larger number of wavelength channels of the microcomb.

Figure 8 shows calculated examples of the estimated outputs of the range gates [43] with different distances. Considering an example with a sequence length $N/m$ = 30, the delay of the matched filters would be $2\Delta t$=168 ps. The tap coefficients for the $l_{th}$ matched filter are $c[k\text{-}l]$, where $c[k]$, $k$=1, 2, …30, is the employed phase codes. The range resolution of the radar, which is the minimum distance between two resolvable targets, is determined by the delay step ($2\Delta t$) of the matched filters, which is given by $2\Delta t \cdot c$=5cm, where $c=3\times10^8$m/s is the speed of RF signals in air. If the distance between the target and radar is $R$, then the delay would be $2R/c$. The range gates would have a maximum output at the $l_{th}$ range gate, $l=2R/(c\cdot 2\Delta t)$. We note here that this calculation only shows the basic connections between our phase encoder and radar systems' performances - practical radar systems are subject to more complicated trade-offs involving capability versus performance.

We note that although random phase codes were used here as a proof of concept, the design of phase codes can be optimized to achieve a better detection performance of the radar system, which is an NP-hard problem and can be approximated and addressed with the methods in [44].

Further, the routes to further scale up the performance of our system are clear. The sequence length can be further increased by either employing microcombs that have a smaller FSR and thus more wavelength channels in the available optical band, or by introducing spatial-multiplexing techniques to boost the length of parallel channels. Moreover, the increasingly mature nonlinear platforms and recent advances in coherent soliton generation techniques (such as deterministic soliton generation [45] and battery-driven solitons [46]) enable microcombs [47-127] to have an increasingly wide access and enhanced performance for engineering applications, including the phase-encoded signal generator reported here.

## IV. Conclusion

We demonstrate photonic RF phase encoding using an integrated micro-comb source. A single-cycle Gaussian pulse was multicast onto the comb wavelengths to assemble the desired phase-encoded RF waveform. A high pulse compression ratio of 29.6 and phase encoding speed of 5.95 Gb/s were achieved, enabled by the use of 60 wavelengths generated by the microcomb. The sequence length was reconfigured by adjusting the length of each phase code, which led to a reconfigurable encoding speed. These results verify that our approach to high-speed RF phase encoding is competitive in terms of performance, with potentially lower cost and footprint.

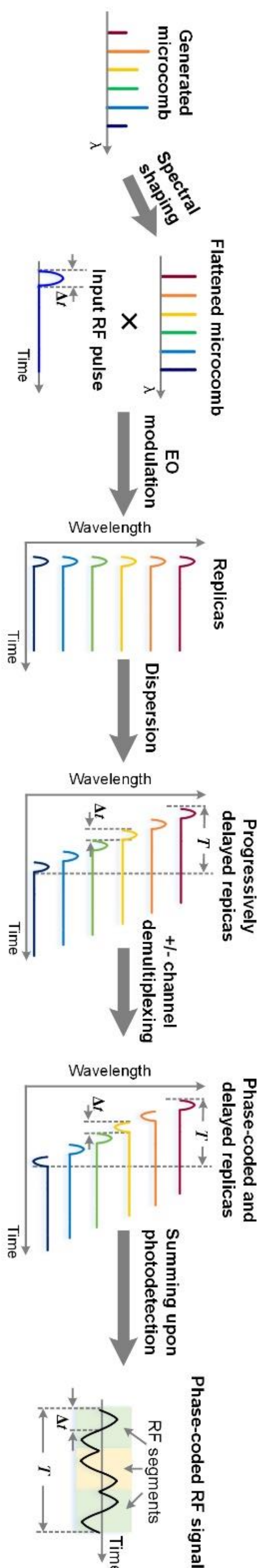

Fig. 1. Illustration of the operation of the photonic RF phase encoder.

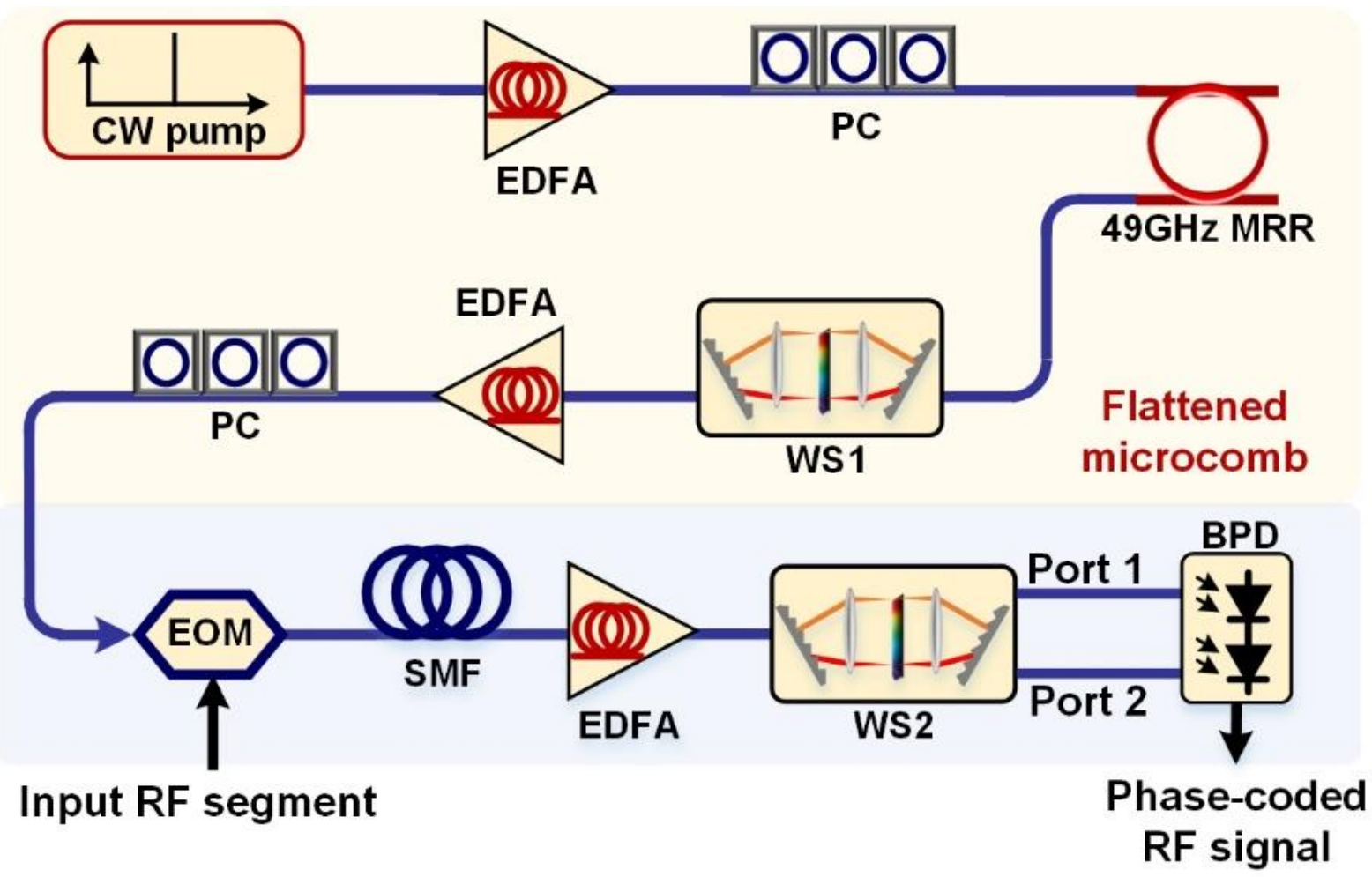

Fig. 2. Schematic diagram of photonic RF phase encoding based on an integrated optical micro-comb source. EDFA: erbium-doped fibre amplifier. PC: polarization controller. MRR: micro-ring resonator. WS: WaveShaper. EOM: Mach-Zehnder modulator. SMF: Single mode fibre. BPD: balanced photodetector.

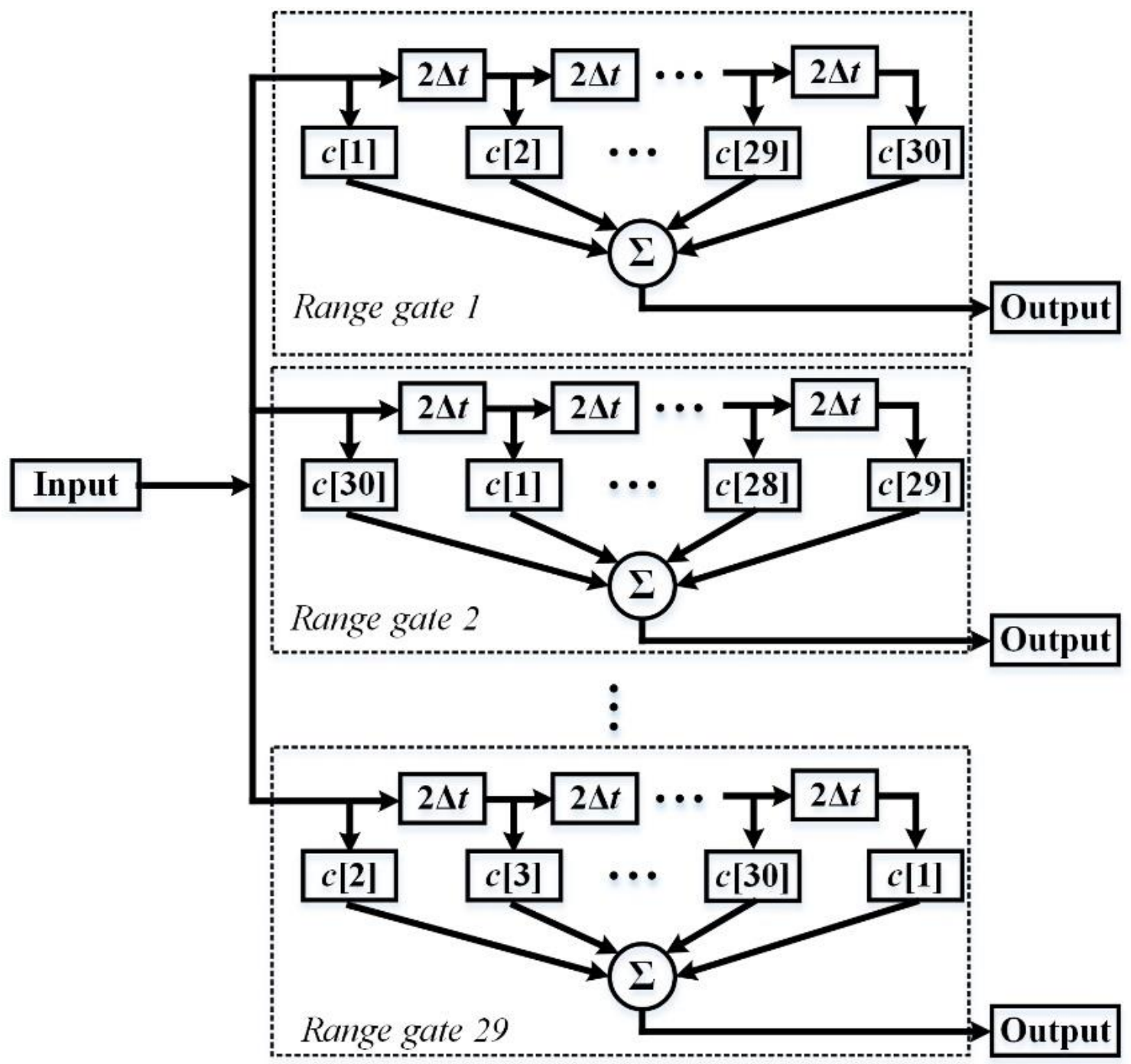

Fig. 3. Structure of the range gates at the radar systems' receiver.

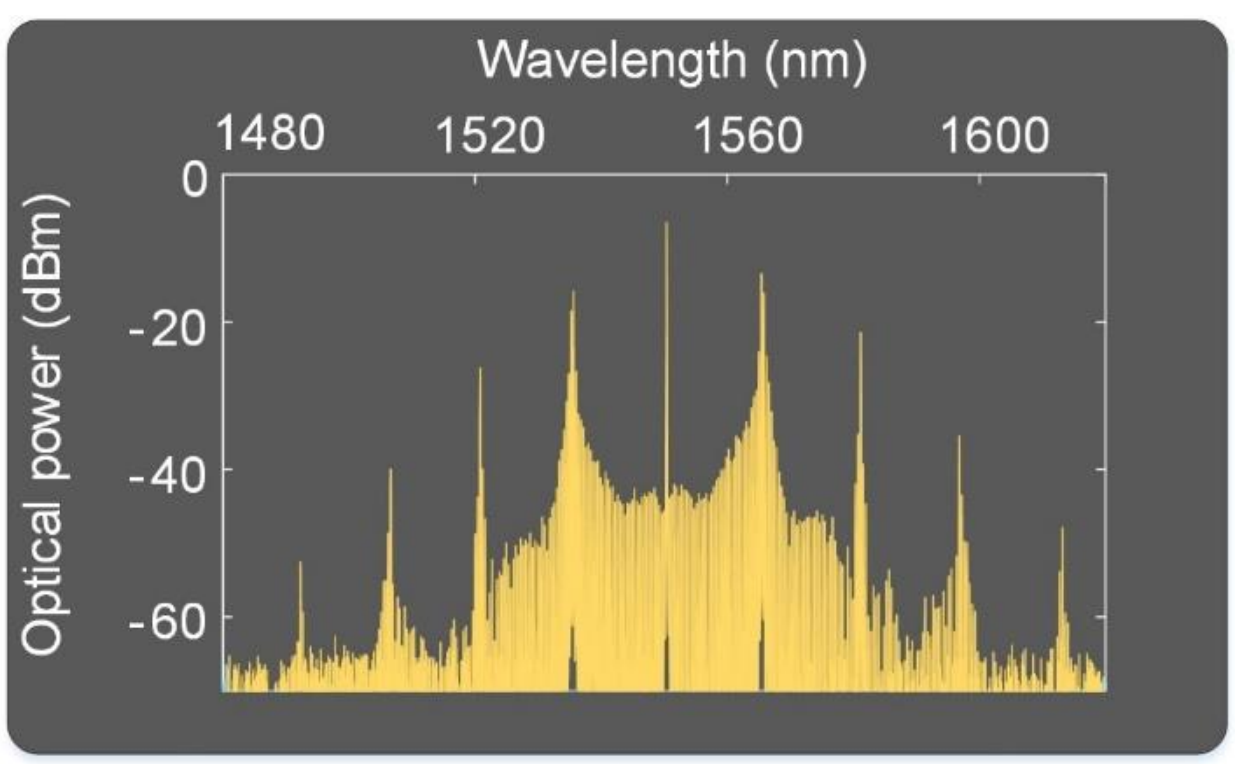

Fig. 4. Soliton crystal microcomb.

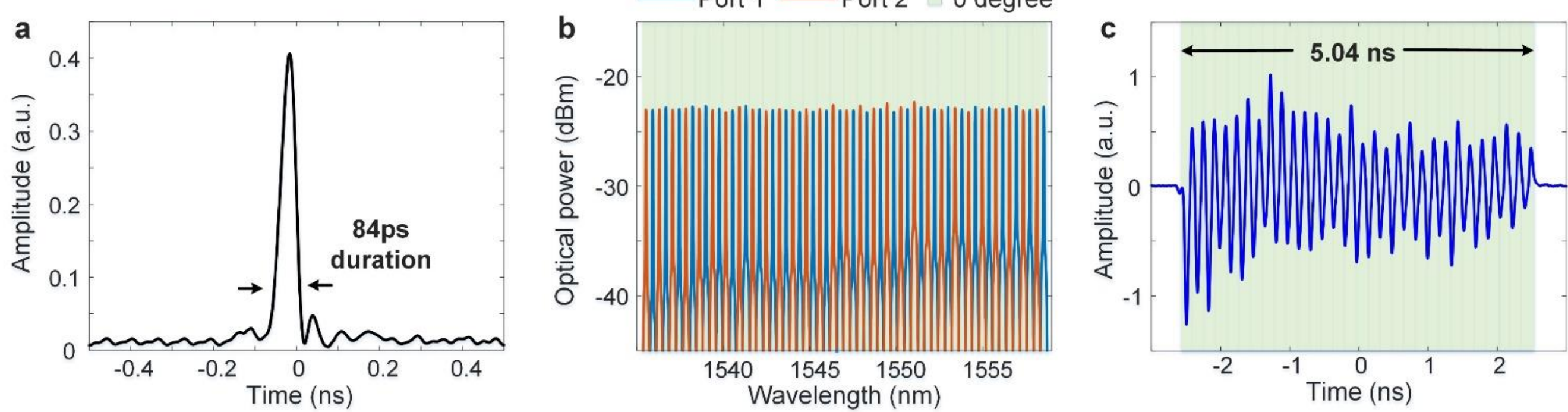

Fig. 5. (a) The input Gaussian pulse. (b) The optical spectra of the flattened microcomb at different ports of the Waveshaper. (c) Assembled RF signal with 30 cosine cycles.

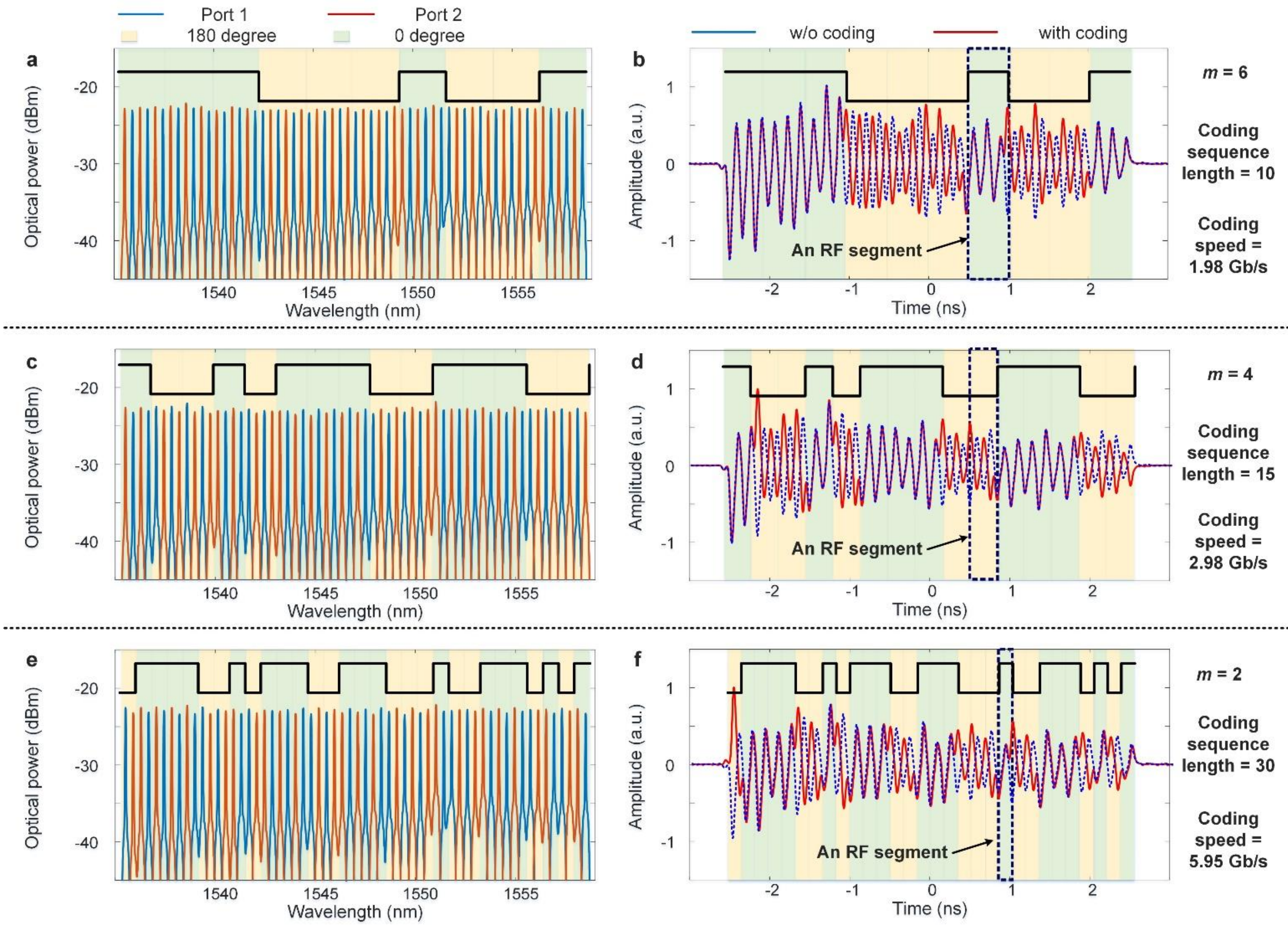

Fig. 6. (a, c, e) The optical spectra of the flattened microcomb at different ports of the Waveshaper with different phase codes and sequence length. (b, d, f) The assembled phase-encoded RF waveform.

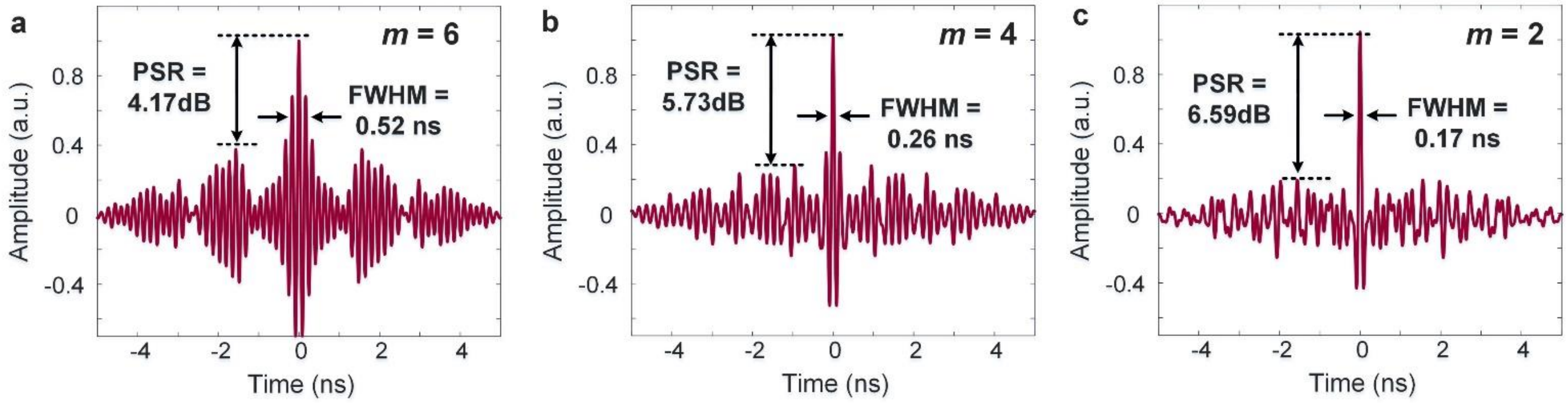

Fig. 7. The calculated autocorrelation of the phase-encoded RF waveforms for (a) $m$=6, (b) $m$=4, and (c) $m$=2.

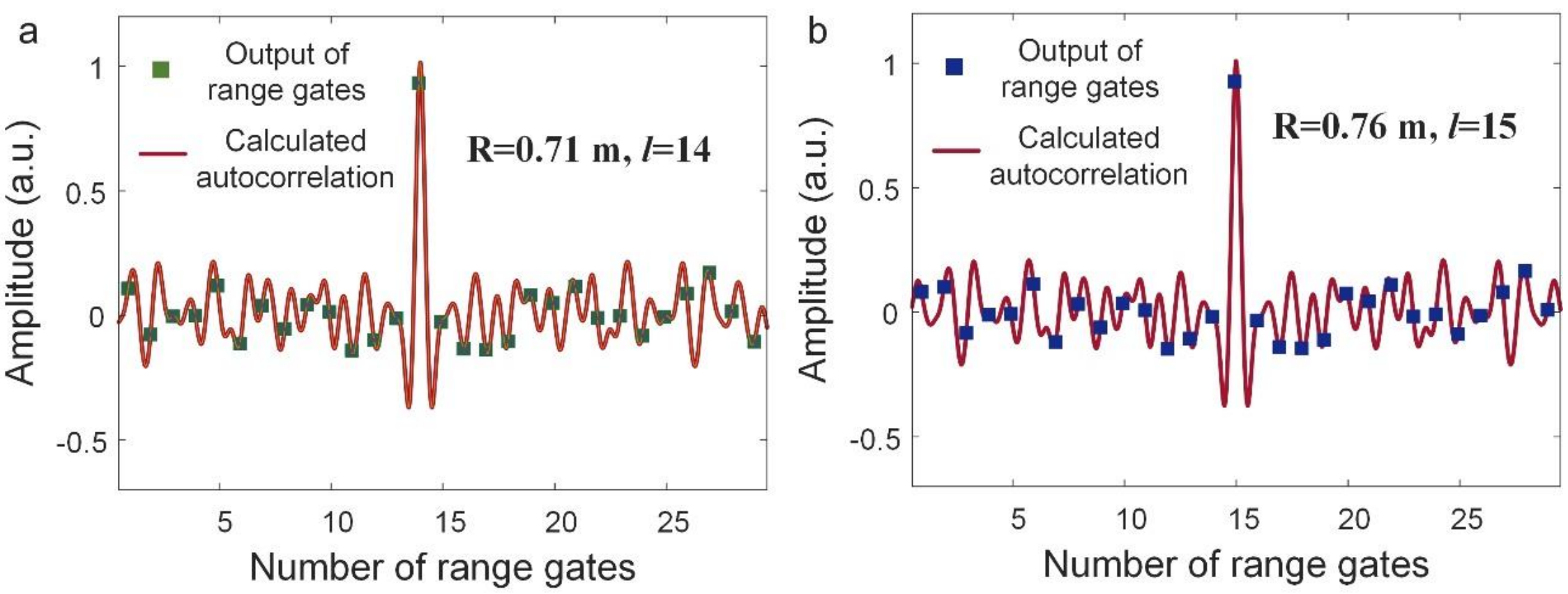

Fig. 8. Calculated outputs of the range gates for (a) $R$=0.71 m and (b) $R$=0.76 m.